# The *NOVO* project: neutron detection for real-time range verification in proton therapy – A Monte Carlo feasibility study

Kristian Smeland Ytre-Hauge[1*], Kyrre Skjerdal[2], John Mattingly[3] and Ilker Meric[3,4]

[1] Department of Physics and Technology, University of Bergen, P.O. Box 7803, 5020 Bergen, Norway
[2] Department of Computing, Mathematics and Physics, Western Norway University of Applied Sciences, P.O. Box 7030, 5020 Bergen, Norway
[3] Department of Nuclear Engineering, North Carolina State University, Raleigh, NC 27695 – 7909, USA
[4] Department of Electrical Engineering, Western Norway University of Applied Sciences, P.O. Box 7030, 5020 Bergen, Norway

**Corresponding author:** Kristian Smeland Ytre-Hauge, kristian.ytre-hauge@uib.no

## Abstract
Uncertainties in the proton range in tissue during proton therapy limit the precision in treatment delivery. These uncertainties result in expanded treatment margins, thereby increasing radiation dose to healthy tissue. Real-time range verification techniques aim to reduce these uncertainties in order to take full advantage of the finite range of the primary protons. In this paper, we propose a novel concept for real-time range verification based on detection of secondary neutrons produced in nuclear interactions during proton therapy. The proposed detector concept is simple; consisting of a hydrogen-rich converter material followed by two charged particle tracking detectors, mimicking a proton recoil telescopic arrangement. Neutrons incident on the converter material are converted into protons through elastic and inelastic (n,p) interactions. The protons are subsequently detected in the tracking detectors. The information on the direction and position of these protons is then utilized in a new reconstruction algorithm to estimate the depth distribution of neutron production by the proton beam, which in turn is correlated with the primary proton range. In this paper, we present the results of a Monte Carlo feasibility study and show that the proposed concept could be used for real-time range verification with millimetric precision in proton therapy.

## Introduction

In cancer treatment, growing numbers of patients are treated with proton therapy. The main rationale for proton therapy is the inverse dose distribution with depth and the finite range of protons which provide dosimetric advantages compared to conventional radiotherapy with photons. The proton range depends on the initial energy of the protons and physical properties of the target material and is normally calculated from a CT scan of the patient.[1] Uncertainties related to the conversion from Hounsfield units in the CT scan to proton stopping power (or range), as well as uncertainties related to tissue heterogeneities, and patient motion and positioning, contribute to the overall uncertainty in the proton range during treatment[2,3]. To assure that treatment goals are achieved, a distal margin (e.g. 3.5% of range + 1 mm or fixed 5 mm)[2] is introduced in clinical protocols to account for the uncertainties in range. If the range uncertainties were reduced, a smaller margin could be applied, thereby sparing healthy tissue. This could also allow for more ideal irradiation field arrangements when organs at risk (OARs) are in close proximity to the target volume. With this motivation, several strategies have been explored in order to monitor and verify the correct delivery of the treatment through measurements during irradiation. These strategies include positron emission tomography (PET) imaging, prompt gamma (PG) imaging, imaging of large angle scattered protons and, for carbon ion therapy, imaging of secondary ions.[4-9] Both PG- and PET imaging have been applied in clinical proton therapy,[10,11] but limitations are still present in these techniques. PET imaging has been primarily performed off-line (after treatment) due to challenges including neutron contamination and limitations in statistics achievable during treatment.[4,12,13] On the other hand, PG imaging has been shown to be a promising tool for real-time range verification, although with challenges resulting from sensitivity to detector positioning, statistical uncertainties and contamination from secondary neutrons degrading the PG detection sensitivities.[2,14] Also, the state-of-the-art in PG imaging is based on mechanically collimated detector arrays leading to a large footprint for such systems.[10]

The principle of detecting secondary radiation produced by the primary beam through nuclear interactions is common to the aforementioned techniques. In addition to positron emitters, photons and secondary ions, secondary neutrons are also produced in interactions along the primary proton beam path. In a recent study, Marafini et al.[15] explored the potential of detecting these neutrons in order to estimate the neutron dose distribution in the patient. Similar approaches are also under exploration e.g. by Lyoussi et al.[16] and Clarke et al.[17], and a novel neutron spectrometer with comparable properties was recently developed by Langford et al.[18] The application of the neutron signal for primary beam range verification has been proposed,[15] but remains an as-yet unexplored range verification modality.

In this work, we propose and investigate the feasibility of a novel and previously unexplored concept for range verification in proton therapy. The NOVO (NeutrOn detection for real-time range VerificatiOn) project is based on the detection of secondary neutrons produced in nuclear interactions during proton therapy for the purpose of real-time proton range verification. Here, we present the results of FLUKA[19,20] (version 2011.2c.6) Monte Carlo (MC) simulations exploring the potential of a simple neutron detection concept consisting of a hydrogen-rich material, converting neutrons to protons mainly through elastic scatters on hydrogen-1 nuclei, followed by two charged particle tracking detectors, mimicking a proton recoil telescopic arrangement. MC simulations of proton pencil beams incident on a homogeneous water phantom, followed by subsequent detection of secondary neutrons forms the basis of the current study. Secondary neutrons incident on the converter material produce protons through elastic scatters on hydrogen-1 nuclei and inelastic interactions, and the two tracking detectors are used to determine the direction and position of these protons. The information on the direction and position of detected protons is used in conjunction with a new reconstruction algorithm based on an MC acceptance/rejection sampling scheme to estimate the neutron production depth distribution in the water phantom which finally can be used for proton beam range estimation.

## Results

**Neutron production characteristics in the water phantom**
The neutron production as a function of depth was seen to be relatively stable in the entrance region of the phantom followed by a steep decrease just proximal to the Bragg peak (Figure 1a). An increase in neutron production rate was also observed with increasing primary proton beam energy, as seen in Figure 1. The rates of neutrons produced in the water phantom per primary proton were 0.11, 0.17 and 0.22 for the primary proton beam energies of 160, 200 and 230 MeV respectively, all with a statistical uncertainty below 1%. The neutron energy spectra (Figure 1b) were dominated by neutrons of energies above 1-10 MeV for all primary proton beam energies and, as expected, the maximum neutron energy increased with energy of the primary beam. In Figure 1c, it can be observed that neutrons produced in the water phantom were predominantly emitted in the forward direction, while the angular distributions were symmetric in the directions perpendicular to the proton beam. As seen in Figure 2a, the neutrons were primarily produced along the primary proton beam axis. It is also observed that neutrons of higher energies were produced in the entrance region of the phantom compared to larger depths.

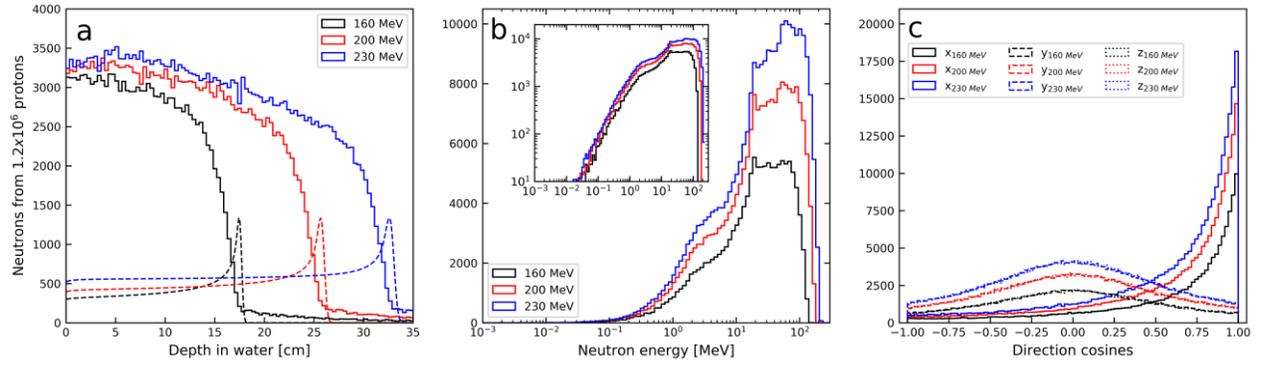

Figure 1: Characteristics of the neutron production in the water phantom for the 160, 200 and 230 MeV proton beams as obtained from MC simulations. a: Neutron production in the water phantom as a function of depth in the phantom. Relative depth doses are shown with dashed lines (arbitrary dose units). b: Distribution of initial kinetic energy of neutrons generated in the water phantom. The neutron spectra are shown with logarithmic bins and with linear (main plot) and logarithmic (inlay) ordinate axis. c: Angular distribution of neutrons produced in the water phantom. Direction cosines are shown for the direction along the beam axis (x, solid histogram) and the y- and z-axes (overlapping dashed and dotted histograms, respectively), perpendicular to the initial beam direction.

**Neutron detection in tracking detectors**

Figure 3a shows the production depth distributions of neutrons that were converted to protons in the converter layer through (n,p) elastic or inelastic collisions and subsequently detected in the tracking detectors in the MC simulation. Compared to Figure 1a, we can observe that the distal fall-off in the distributions is shifted towards shallower depths and exhibits less steep fall-off. For ten repeated simulations with $1.2 \times 10^9$ protons, the average rates of detected neutrons in the tracking detectors per primary proton were $2.0 \times 10^{-5}$ (±0.5%), $3.4 \times 10^{-5}$ (±0.2%) and $4.2 \times 10^{-5}$ (±0.5%) for the primary proton beam energies of 160, 200 and 230 MeV, respectively. The energy distributions of the detected neutrons are shown in Figure 3b and ranges from 20 MeV and almost up to the primary beam energy. The cut off at 20 MeV was caused by the exclusion of protons from *low energy neutron scattering interactions* ($E_n < 20$ MeV), as described in the methods section. The angular distribution of detected neutrons (Figure 3c) is slightly broadened, i.e. less forward-peaked, compared to the angular distribution of all neutrons produced in the water phantom (Figure 1c). Also, due to the limited solid angle subtended by the detector, the ranges of direction cosines for the y and z-axes are reduced for the detected neutrons compared to direction cosines observed for all neutrons. Neutron production positions and energies for detected neutrons are shown in Figure 2b. Comparing to the results in Figure 2a, it can again be observed that the detected neutrons overall have higher energies, in particular the detected neutrons originating from the entrance region. This is again partly due to the exclusion of data from *low energy neutron scattering interactions*. Furthermore, the data presented in Figure 2b reveals a reduction in the neutron detection rate close to the Bragg peak for all energies.

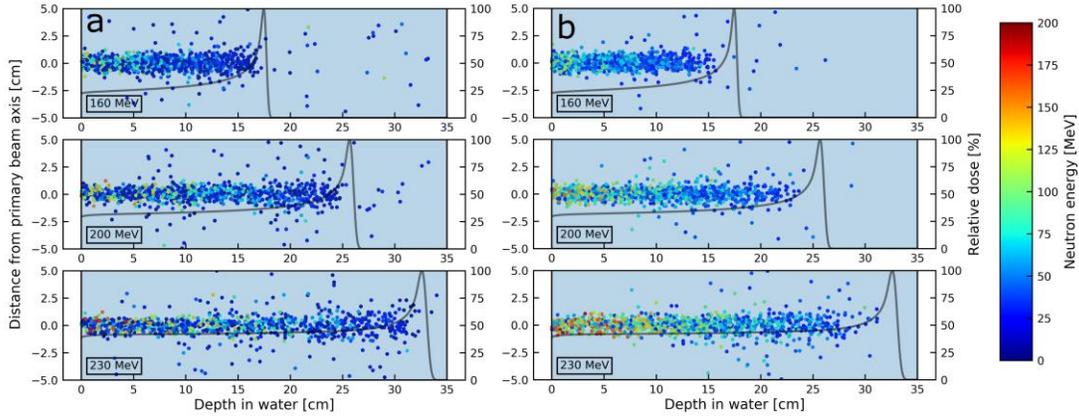

Figure 2: Production positions (shown for 1000 neutrons) inside the water phantom for a set of arbitrarily chosen neutrons (panel a) and detected neutrons (panel b). The kinetic energies of the neutrons are indicated by the colorbar. Relative depth doses are shown on the right y-axis.

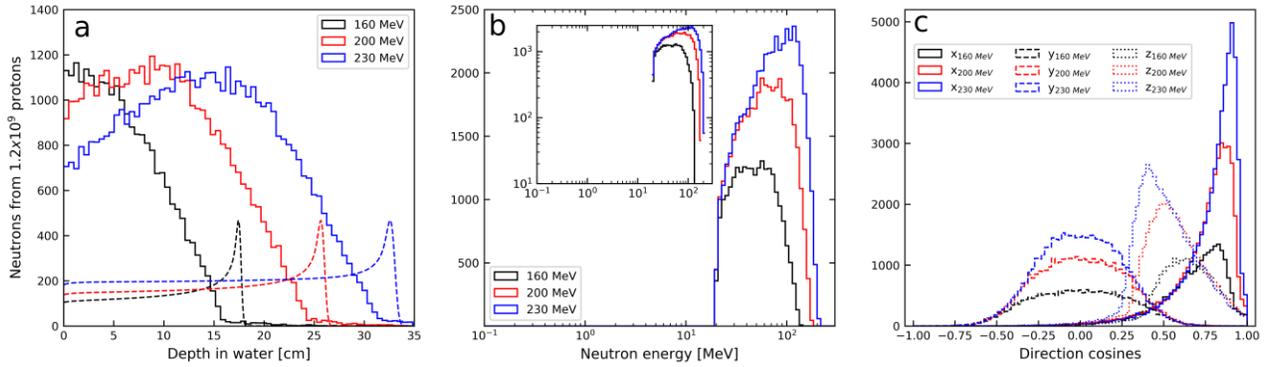

Figure 3: Characteristics of the detected neutrons produced in the water phantom for the 160, 200 and 230 MeV proton beams as obtained from MC simulations. a: Production depth distributions for detected neutrons. Relative depth doses are shown with dashed lines (arbitrary dose units). b: Distribution of initial kinetic energy of detected neutrons. The neutron spectra are shown both with linear (main plot) and logarithmic (inlay) ordinate axis. The absence of data below 20 MeV is due to the exclusion of low energy neutron scattering events ($E_n < 20$ MeV) in the reconstruction of neutron production distributions. c: The initial angular distribution of detected neutrons. Direction cosines are shown for the direction along the beam axis (x, solid histogram) and the y- and z-axes (dashed and dotted histograms), perpendicular to the initial beam direction.

**Reconstructed neutron production depth distributions**

The neutron production distribution in the homogeneous water phantom reconstructed using the information from the tracking detectors are shown in Figure 4 for the three primary proton beam energies considered in this work. The reconstructed distributions (solid lines in Figure 4) reveal lower rates at shallow depths for all three energies, compared to the MC ground-truth, i.e. the true distributions from MC simulations (dotted lines in Figure 4). For the two highest proton energies, lower rates were also observed for the reconstructed distributions close to the Bragg peak, while in the intermediate region the rate is higher for the reconstructed distributions. Comparison of distributions from MC ground-truth and reconstruction gave RMSE (root mean square error) values of approximately 10% or below, as also indicated in Figure 4. Overall, the trend in the distal fall-off of the reconstructed distributions is similar to that of the ground-truth from MC simulations dropping to zero proximal to the Bragg-peak.

Using the reconstruction algorithm (for a total of 5000 iterations per detected proton) on data from $1.2 \times 10^9$ primary protons, the range landmark positions, β, of the reconstructed distributions were estimated to 10.62±0.04 cm, 14.61±0.03 cm and 18.76±0.03 cm for the 160, 200 and 230 MeV proton beams, respectively. The corresponding landmarks obtained from MC ground truth data were 10.04±0.04 cm, 15.94±0.02 cm and 21.49±0.04 cm. Uncertainties in the landmark determination as a function of primary beam intensity are shown in Figure 5. For all primary proton beam energies, the uncertainties in the range landmark position remained below 2 mm for the reconstructed landmarks and below 2.3 mm for the landmarks from MC ground-truth for primary proton beam intensities as low as $5 \times 10^7$. The uncertainties shown in Figure 5 reveal negligible dependence of the uncertainty on the primary proton beam energy whereas the dependence on the primary proton beam intensity is evident. Further, the uncertainties in β-values from the reconstructed and MC ground-truth data are similar in magnitude, indicating that the reconstruction algorithm does not introduce additional uncertainties in the estimates of the range landmark. The correlation between β and the primary beam range is shown in Figure 6, both for reconstructed- and MC ground-truth. The calculated β-values indicate an almost linear correlation with the primary proton beam range as depicted in Figure 6 through linear fits to the data, and the resulting $R^2$ values. The data given in Figure 6 are based on $1.2 \times 10^9$ primary protons for each energy, thus with sub-millimetre uncertainties, as reported in Figure 5 and indicated by the error bars in Figure 6.

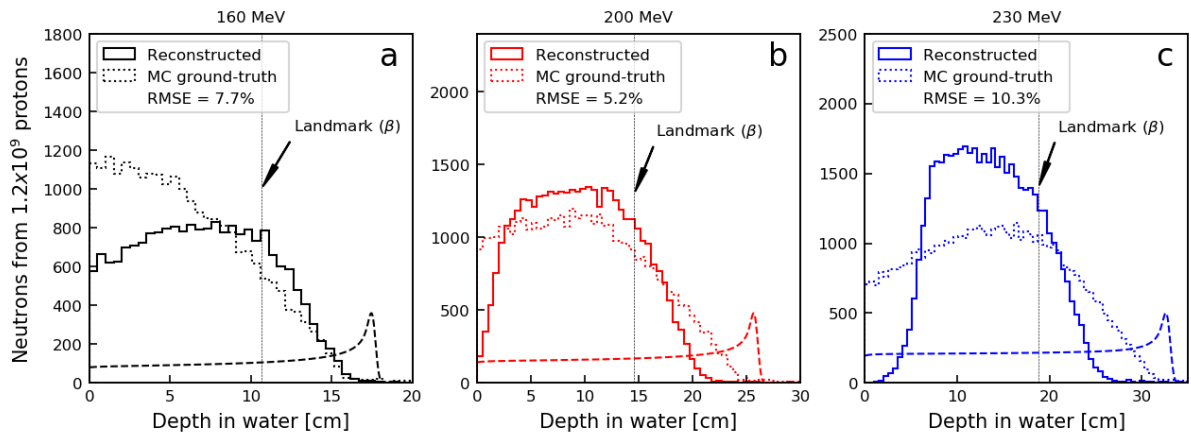

Figure 4: Distribution of production depths for detected neutrons produced in 160 (a), 200 (b) and 230 (c) MeV proton beams. Both the ground-truth distributions obtained directly from MC simulations (dotted lines) and the reconstructed profiles (solid lines) are shown. The β landmark for range verification, obtained from reconstructed distributions, is indicated on all figures with vertical dashed lines.

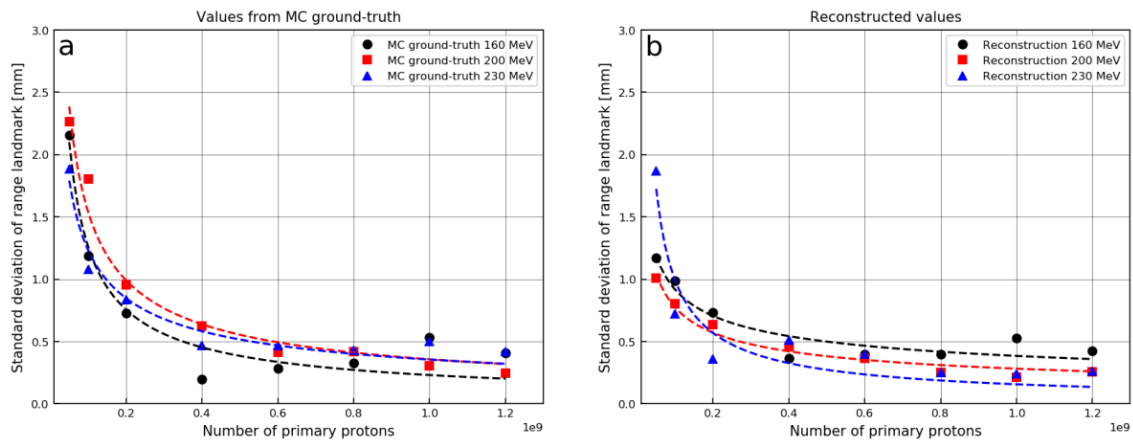

Figure 5: Uncertainties in range landmark determination as a function of number of primary protons for 160, 200 and 230 MeV proton beams shown both for MC ground-truth (a) and reconstructed values (b). In the reconstructions, a total of 5000 iterations per detected proton was performed. Uncertainties for both MC ground-truth and reconstructed data are calculated from ten independent repetitions for each energy and for each primary proton beam intensity.

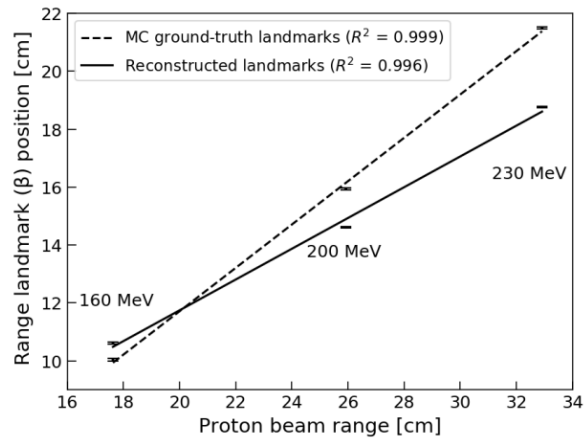

Figure 6: The plot shows the landmark positions (β) obtained from the reconstruction algorithm as a function of the proton range. A linear curve has been fitted to the data, describing the energy dependence of the landmark position. Also given is the $R^2$ value of the linear fit.

## Discussion and Conclusion

In this study, FLUKA MC simulations were used to explore the potential for real-time range verification in proton therapy based on detection of secondary neutrons. A setup with a homogeneous water phantom and the proposed detector concept was implemented in the MC simulations. The neutron production rate and the energy distribution of the secondary neutrons produced in the water phantom were observed to depend strongly on both the depth in the phantom as well as primary proton beam energy. Higher energy neutrons were produced in the entrance region where the primary proton beam energy is still high. Secondary neutrons were, with few exceptions, produced close to the primary beam axis. These results are in good agreement with previous MC studies.[21] Clear differences were observed between the characteristics of the detected neutrons (Figure 3) compared to all neutrons produced in the water phantom (Figure 1). In particular, the angular distribution of detected neutrons deviated from the overall angular distribution of neutrons produced in the water phantoms. This implies that the positioning of the detector may influence the production depth distributions of detected secondary neutrons. In this work, the detector was centred at Bragg peak depth to balance the need for detecting neutrons both prior to and in the Bragg peak region. The importance of detector positioning has previously been discussed for PG imaging both in terms of distance to the isocentre and the detector acceptance angle,[22,23] and a detailed study of the impact of detector positioning could potentially lead to a more favourable detector placement.

The neutron detection rates for the current design, determined to be approximately 2-4x$10^{-5}$ neutron counts per primary proton, are comparable to detection rates obtained with existing PG imaging systems, i.e. $10^{-5}$ – 5.6x$10^{-5}$ gamma counts per primary proton, reported in the literature[5]. Still, a detailed detector design study aiming at increasing the detection efficiency could further reduce the statistical uncertainties in the range estimates. This will be of importance at beam intensities lower than about 5x$10^7$ protons (Figure 5). Polf et al.[24] suggested that range verification systems should be capable of resolving range shifts with good precision at primary proton beam intensities as low as 4.5x$10^7$. However, as the number of protons in a clinical treatment plan varies with the target size, range verification on a spot-by-spot basis may require imaging also at lower primary beam intensities. The fraction of neutrons reaching the detector is inherently limited by the solid angle it covers, and it could be increased e.g. by increasing the detector area. As neutrons are predominantly emitted in the forward direction, placing the detector at angles < 90° with respect to the incident proton beam direction could increase the number of detected neutrons as well as the quality (in terms of the RMSE-values) of the reconstructed production position distributions. This is because neutrons would, on average, hit the converter at smaller angles with respect to the converter surface normal. It should also be noted, although not considered in this feasibility study, there will be an inevitable contamination from large angle scattered protons which need to be eliminated. Thus, there is likely a need to shield the detector system from large angle scattered protons originating from the patient or phantom, and this should be evaluated thoroughly in MC studies before a first prototype detector is constructed.

The comparison of MC ground-truth distributions and the reconstructed distributions (Figure 4) indicates that there is room for improvement also in the reconstruction algorithm. In general, the algorithm induces a compression of the depth production distributions, especially observed for high primary proton beam energies. This is most likely because sampled neutron incidence angles are accepted or rejected based on probability density functions (PDFs) that describe the angular distributions of incident neutrons along the primary proton beam axis (see Figure 2c). These distributions are forward-peaked and thus result in a high rejection rate at the tails of the distributions. Adjustments in the acceptance/rejection regime could lead to reduced RMSE values in the comparison of MC ground-truth and reconstructed distributions. However, it must be noted that, while possible improvements of the algorithm should be further explored, a better fit to MC ground-truth data does not guarantee improved repeatability of the landmark determination or reduction in the uncertainty in the range estimates.

As the analytical shapes of the resulting neutron production distributions (for both reconstructed and MC ground-truth distributions) are not known, we decided to use instead of the full distribution, the sum of the sample mean for the ensemble of reconstructed coordinates and the corresponding standard deviation as an estimate of the range landmark (step 10 of the reconstruction algorithm). This simple means of calculating the range landmark turned out to be sufficiently robust for the purposes of this feasibility study, giving relatively small standard deviations in the range landmark estimates as shown in Figure 5, and a linear relation between the range landmark and the primary proton beam range as shown in Figure 6. Further, while the uncertainties in the landmark determinations could be seen to depend on the primary beam intensity (Figure 5), an energy dependence, due to the lower neutron production could also be expected. Most probably, this will be more prominent at even lower beam intensities, where the statistics available for reconstruction of production distributions will be more limited.

In this work, we focused on the one-dimensional (1D) distributions of neutron production positions in a water phantom and have shown, for the first time, that the information available through 1D distributions would be sufficient for sub-millimetric determination of a range landmark, β, that can be correlated to the primary proton beam range. However, the information from the tracking detectors could, in principle, be used also for 2D or 3D reconstruction of the neutron production distributions. 2D and 3D distributions of neutron production positions could potentially be utilized for a more comprehensive dose verification or dose reconstruction technique. Such information can also be used for determination of the neutron dose component which, as of today, is normally not considered by commercial treatment planning systems.

The results presented in this work indicate that the reconstructed neutron production position distributions and the estimated range landmark positions are dependent on, and can be correlated to, the primary proton beam range. From both MC ground-truth and reconstructed distributions, it is observed that the standard deviation in the reconstructed range landmark positions is at about 2 mm or less for all three energies at proton intensities as low as $5 \times 10^7$. The resulting uncertainties are similar to or below the uncertainties that are reported in the literature for the state-of-the-art PG imaging systems[14,25] based on MC simulations of these detector systems. It is also anticipated that the resulting uncertainties in the range landmark estimates can be further reduced through increasing the detector acceptance and area which, in turn, will result in increased neutron detection rates per primary proton. In light of these considerations as well as the results presented in this work, we conclude that the proposed secondary neutron detector concept is a promising alternative for range verification in proton therapy with a statistical precision of 1-2 mm, in agreement with the general consensus on the required precision of such systems.[26]

# Methods

### Conceptual design and detection principles

The basis for this study of range verification in proton therapy using neutrons was a MC simulation with a water phantom (35x10x10 cm$^3$) irradiated by proton pencil beams of energies 160, 200 and 230 MeV and spatial Gaussian profiles of 10 mm full width at half maximum. MC simulations were performed with the FLUKA code, version 2011.2c.6. The range verification system was implemented with three components, as seen in Figure 7; a 5 mm thick converter layer made from the hydrogen rich organic scintillator material EJ309 (atomic ratio H/C:1.25, density 0.959 g/cm$^3$), followed by two proton tracking detectors, covering an area of 20x20 cm$^2$. The detector was centred at the Bragg peak depth, 10 cm lateral to Bragg peak position. Neutron detection was based on (n,p) interactions in the converter material, and subsequent detection of the protons in both tracking detectors. Using positional information from the two detectors, the direction of the proton could be determined. The proton vectors obtained using the position information from the two tracking detectors formed the basis for estimating the distribution of neutron production in the water phantom. Finally, range landmarks (β) calculated from the estimated neutron distributions were correlated to the range of the primary proton beam.

### Monte Carlo simulations

The MC simulations were used to obtain detailed information about neutron production in the water phantom and the propagation and detection of these neutrons in the proposed detector system. In addition, depth dose profiles and beam range (defined as the depth of distal 80% dose) estimates were obtained. Tracking of secondary neutrons and neutron induced protons from the converter layer were implemented in FLUKA: All neutrons produced in the water phantom through nuclear interactions were flagged and their position of origin, direction and energy were saved during the simulations. Protons produced in inelastic and elastic (n,p) reactions in the converter material were set up to inherit the information about their *parent* neutrons. Thus, upon registration of the protons (produced in the converter) in the tracking detectors, the properties of the parent neutrons could be extracted simultaneously. Protons from the converter were also classified in three (FLUKA specific) categories depending on the type of interaction they were created in: 1) Inelastic interaction, 2) Elastic interaction and 3) Low energy neutron scattering ($E_n$ < 20 MeV). For the latter category, the data were not included in the further analysis as the low proton energies (and ranges) from these interactions will result in lower probabilities of successful detection in two tracking detectors in a realistic detection system. The positional information from protons of categories 1) and 2) in the two tracking detectors were used for reconstructing the neutron production depth distributions. The additional information on the parent neutron available in the MC simulations enabled evaluation of the reconstruction method.

The MC simulations were performed with 1.2x10$^9$ (5x10$^7$ x 24 CPUs) primary proton histories. The simulations were repeated ten times to allow statistical analysis of the results. As a single run was distributed across 24 CPUs each running 5x10$^7$ protons, the simulation results were divided into 24 bunches. To obtain uncertainties at e.g. proton intensity of 1.2x10$^9$ all 24 bunches were included in the analysis. The resulting uncertainties in the range landmark estimates ($\sigma_\beta$), were then calculated as:

$$\sigma_\beta = \sqrt{1/9 \sum_{i=1}^{10}(\beta_i - \bar{\beta})^2}, \tag{1}$$

Where $\beta_i$ is the landmark for repetition i, and $\bar{\beta}$ is the mean landmark value of the ten repeated simulations. Next, leaving out four bunches in each of the ten repetitions allowed calculation of the uncertainties at proton intensity of $1.0 \times 10^9$. Further reductions in the number of bunches included in the analysis then allowed estimations of the uncertainties in the calculated range landmark positions, β, for both the MC ground-truth and reconstructed profiles as a function of the number of primary proton histories.

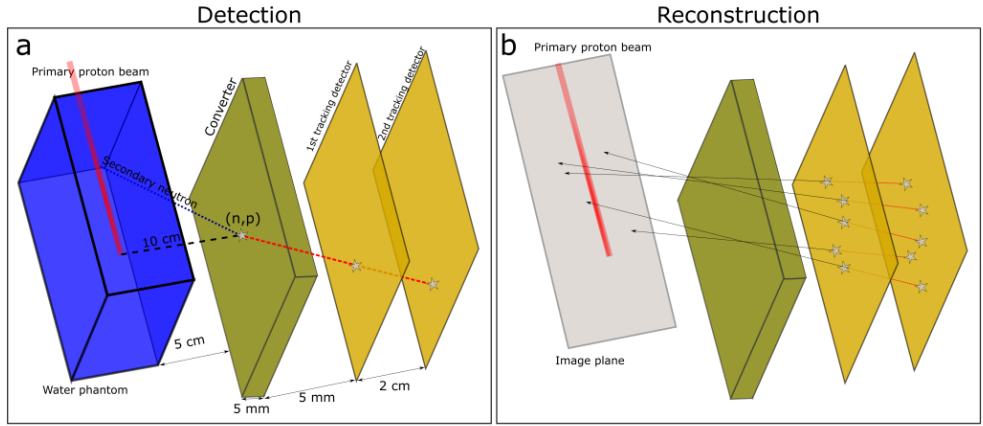

Figure 7: The conceptual design (a) as implemented in the Monte Carlo simulations (not to scale). Neutrons produced along the beam path in the water phantom may be converted to protons in the converter material. Protons reaching both tracking detectors give positional information for reconstruction (b) of the production point of the parent neutron.

**Reconstruction and range landmark determination**

The aim of the reconstruction algorithm is to determine a reproducible range landmark, β, which can be correlated to the primary proton beam range. For this purpose, we developed an iterative MC acceptance-rejection sampling scheme, utilizing the information available from the tracking detectors in the MC simulations through back projection of the detected proton vectors onto a plane along the initial primary beam, hereafter referred to as the *image plane*. It should be noted that imaging of neutrons traditionally relies on reconstructing and back projecting an event cone following double elastic (n,p) scatterings in the detector material[27]. This is not the case in our proposed design as we rely on measurements of the direction of a proton from a single (n,p) event. Also, we take into account not only elastic scatter events but also inelastic scatters. Thus, the reconstruction problem encountered in this study must be considered an ill-posed inverse problem. To partially mitigate this, a MC sampling approach was considered where we also incorporate prior information about the water phantom and the angular distributions of detected neutrons in addition to using the initial information available through back projection of the vectors of protons detected in the two tracking detectors.

The reconstruction algorithm is based on the following steps:

1. The spatial information from proton detection in the two tracking detectors is used to construct vectors for back projection of the proton path onto an *image plane* along the initial primary beam axis. The *image plane* crosses the centre of the primary proton beam along the lateral direction as shown in Figure 7.
2. The initial back projection onto the image plane gives coordinates that are scattered throughout the image plane. We then calculate the centre-of-gravity, or unweighted mean ($\mu_x$, $\mu_y$) of the ensemble of back projected coordinates as well as the corresponding standard deviations ($\sigma_x$, $\sigma_y$).
3. For each back projected coordinate we calculate the corresponding Euclidean distance, weighted by the calculated standard deviations along each dimension (essentially the Mahalanobis distance without covariances),[28] to the centre-of-gravity:

$$d_{(x_i,y_i)} = \sqrt{\frac{(x_i-\mu_x)^2}{\sigma_x^2} + \frac{(y_i-\mu_y)^2}{\sigma_y^2}}, \text{ where, } i = 1 \dots N \qquad (2)$$

where $x_i$ and $y_i$ are the $i^{th}$ back projected coordinates, $d_{(x_i,y_i)}$ is the calculated distance from the centre-of-gravity for the $i^{th}$ coordinate and N is the total number of back projected coordinates.

4. We set-up a sampling region that is centred at ($\mu_x$, $\mu_y$) and extends to the maximum distance found in step 3, i.e.: ($\pm \max_i (d_{(x_i,y_i)})$).
5. We sample the polar and azimuthal scattering angles from uniform distributions, θ: (0, π/2) and ϕ: (0, 2π), respectively, and perform a rotation of the corresponding proton vector according to the sampled angles. Following the rotation of the proton vector, the resulting vector is taken as a candidate vector giving the possible incident neutron direction as well as the candidate scattering angle. At this point, we introduce prior information on the incidence angles of neutrons giving an (n,p) collision in the converter material obtained through MC simulations. The sampled, candidate neutron vectors are accepted or rejected based on probability density functions describing the distribution of direction cosines of incident neutrons along the primary proton beam axis. These distributions are highly forward-peaked along the primary beam axis and will thus result in a high rejection rate at the tails of the distributions for all primary beam energies (see Figure 3C).
6. For the accepted angles sampled in the previous step, we back project the corresponding vectors onto the *image plane*. We now perform the next level of acceptance/rejection sampling; we accept the back projected coordinate only if it falls within the pre-defined sampling region, specified in step 4. Here we also incorporate more prior information about the phantom and require that the back projected coordinate falls within a region that is contained within the phantom dimensions in the lateral direction.
7. We repeat steps 5 and 6 for a pre-determined number of iterations, *M*, for each proton, *i*.
8. Once all iterations are completed for a given proton vector, we take the average of all accepted points as an estimate of the production location of the parent neutron.
9. We repeat steps 5 – 8 for all protons from (n,p) collisions that are detected in both tracking detectors and obtain estimates of the production locations of all corresponding parent neutrons.
10. From the resulting distributions, we calculate the range landmark, β, position by first projecting all coordinates onto the axis along primary proton beam and then using the sample mean for the ensemble of reconstructed coordinates ($\mu_{rec}$) and their corresponding standard deviation ($\sigma_{rec}$) as, β = $\mu_{rec.}$ + $\sigma_{rec.}$

For comparison to the reconstruction, the range landmarks were calculated in the same manner also for the true coordinates from MC simulations. Evaluation of the reconstruction algorithm was also performed through RMSE calculations as a measure of the degree of agreement between histograms for the neutron production distribution obtained directly from MC simulations with those obtained through reconstructions. RMSE was calculated for histograms normalized by the highest bin value separately for each of three initial beam energies.

## Data Availability
The datasets generated and analysed during the current study are available from the corresponding authors on reasonable request.

**Acknowledgements**
IM would like to acknowledge the financial support from Bergen Research Foundation (Grant no. BFS2015PAR03).

**Author contributions**
KSYH and IM designed the study, KSYH carried out MC calculations, IM developed the reconstruction algorithm, KSYH, KS, JM and IM interpreted the results, KS and JM provided feedback on the manuscript. KSYH and IM primarily wrote the manuscript with input from all other co-authors.

**Competing interests**
The authors declare no competing interests.